# An Augmented Regression Model for Tensors with Missing Values

Feng Wang, Mostafa Reisi Gahrooei*, Zhen Zhong, Tao Tang, and Jianjun Shi

*Abstract*—Heterogeneous but complementary sources of data provide an unprecedented opportunity for developing accurate statistical models of systems. Although the existing methods have shown promising results, they are mostly applicable to situations where the system output is measured in its complete form. In reality, however, it may not be feasible to obtain the complete output measurement of a system, which results in observations that contain missing values. This paper introduces a general framework that integrates tensor regression with tensor completion and proposes an efficient optimization framework that alternates between two steps for parameter estimation. Through multiple simulations and a case study, we evaluate the performance of the proposed method. The results indicate the superiority of the proposed method in comparison to a benchmark.

*Note to Practitioners*—The proposed method aims to obtain an accurate estimation of the regression model when certain entries of the response are inaccessible. By considering both the information from multiple inputs and the structure of the response, our proposed method can achieve more accurate estimation of the output tensor. In order to apply the proposed method in practice, two assumptions should hold: First, the response tensor should be low-rank, meaning that fewer variation patterns should exist in the response than its dimensions. Second, the relationship between the input tensors and the response should be linear or approximately linear. The presented method in this paper uses tensor decomposition techniques to exploit the correlation structures of the high-dimensional data and prevent overfitting. Another benefit of our integrated framework is that the rank of the response tensor converges automatically, which can be used directly in the parameter estimation.

## NOMENCLATURE

| | |
|---|---|
| Lower- and upper-case letter | A scalar, e.g., $w$ or $W$. |
| Lower- or upper-case boldface letter | A vector or a matrix, e.g., $\mathbf{w}$ or $\mathbf{W}$. |
| Euler script letters | A tensor, e.g., $\mathcal{W}$. |
| $\mathcal{X}_j$ | Input tensor $j$ for the regression model, $\mathcal{X}_j \in \mathbb{R}^{I \times P_{j1} \times \cdots \times P_{jI_j}}$ with the sample dimension $I$ and other dimensions $P_{jk}$, $j = 1, \cdots, p$. |
| $\mathcal{Y}$ | Output tensor for the regression model, $\mathcal{Y}_i \in \mathbb{R}^{I \times Q_1 \times Q_2 \times \cdots \times Q_d}$. |
| $\mathcal{B}_j$ | Coefficient tensor $j$ for the regression model with respect to $\mathcal{X}_j$. |
| $\mathcal{C}_j$ | Core tensor of Tucker decomposition to $\mathcal{B}_j$. |
| $W_{(k)}$ | Mode-$k$ matricization of tensor $\mathcal{W}$. |
| $\Omega$ | Index set to define the missingness of the response. |
| $\mathcal{P}_\Omega$ | Orthogonal projection based on the index set $\Omega$. |
| rank$(\cdot)$ | Rank function of a tensor. |
| $\mathcal{Y}_0$ | Response with missing values. |
| $\mathbf{U}_{j\cdot}$ | Bases that spans the space of the input tensor $j$. |
| $\mathbf{V}_{\cdot}$ | Bases that spans the space of the response. |
| $\mathcal{M}_i$ | The $i^{th}$ local copy of tensor $\mathcal{Y}$. |
| $\mathbf{M}_{(i)}$ | Mode-$i$ matricization of $\mathcal{M}_i$. |
| $\mathbf{\Theta}_i$ | Dual variable. |
| $\lambda$ | Tuning parameter. |

## I. INTRODUCTION

THE rapid development of sensing and computing technology has accelerated the collection of heterogeneous sets of data, which may include a combination of scalars and high-dimensional (HD) data points such as profiles, images, and point clouds. Many applications, including multistage manufacturing [1], aircraft prognosis [2], and medical

(*Corresponding author: Mostafa Reisi Gahrooei.*)

F. Wang and T. Tang are with State Key Laboratory of Rail Traffic Control and Safety, Beijing Jiaotong University, Beijing 100044, China (e-mail: f_wang@bjtu.edu.cn; ttang@bjtu.edu.cn).

M. R. Gahrooei is with the Department of Industrial and Systems Engineering, University of Florida, Gainesville, FL 32611, USA (e-mail: mreisigahrooei@ufl.edu).

Z. Zhong and J. Shi are with the H. Milton Stewart School of Industrial and Systems Engineering, Georgia Institute of Technology, Atlanta, GA 30332, USA (e-mail: zhongzhen@gatech.edu; jianjun.shi@isye.gatech.edu).

TABLE I. REGRESSION METHODS

| Methods | Literature | Characteristics |
|---|---|---|
| Principle component regression (PCR) or partial least square (PLS) | [4], [5] | -Fail to exploit the ordering or spatial structure of profiles or images. |
| Functional regression | [6], [7] | -Mostly on profile data.<br>-Difficult to extend to higher dimensions. |
| Tensor regression (TOT, MTOT) | [12], [13], [14] | -Tensor inputs and tensor output.<br>-Assumes complete data. |

TABLE II. TENSOR COMPLETION METHODS

| Methods | Literature | Characteristics |
|---|---|---|
| Decomposition based methods | CP [18], Tucker [19] decompositions | -With tensor decomposition methods.<br>-Rank specified. |
| Rank minimization based methods | CP [17], Tucker [20] rank | -Automatic rank estimation |

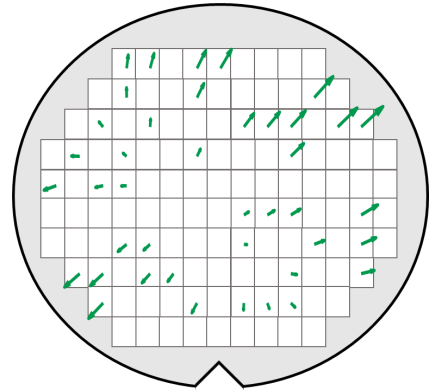

Fig. 1. Illustration of overlay measurements in the lithography process.

imaging [3] can benefit from such heterogeneous and high-dimensional data. For example, by integrating a set of heterogeneous data, including machine settings and sensor readings, a more accurate and reliable prediction of an output of a manufacturing process can be achieved. Such a predictive model can then be used for process monitoring and optimization. As reported in Table I, many techniques have been developed for modeling systems based on HD and heterogeneous data.

Traditional dimensionality reduction approaches, including principal component regression (PCR) [4] and partial least square (PLS) [5] have been used for regression modeling with limited predictive power as they cannot exploit the complex correlation structures of HD data points. Functional regression models are proposed to capture the nonlinear correlation structures of data. Although these methods perform well in cases with profile data, efficient extensions of these approaches to data sets with higher dimensions (e.g., images) are very difficult, if not impossible [6], [7]. In recent years, tensor algebra has shown promising results in many applications, from network analysis to process modeling and monitoring [8], [9]. Among those, some studies have focused on tensor regression analysis. For example, Zhao, *et al.* [10] extended PLS approach to tensor data by using Tucker decomposition. Li, *et al.* [11] proposed a generalized linear model with input tensors based on Tucker decomposition. Yan, *et al.* [12] proposed to link structured point clouds to scalar process variables with tensor regression. To consider a more general case, Lock [13] proposed a regression model that estimates an output tensor based on an input tensor (tensor-on-tensor regression [TOT]), using CP decomposition imposed on the tensors of model parameters. However, the TOT method is limited due to the inherent limitations of CP decomposition and can only include a single input tensor. Recently, Gahrooei, *et al.* [14] extended TOT to multiple tensor-on-tensor regression (MTOT) by leveraging Tucker operations.

Although these existing approaches provide effective ways to model processes using HD data, they assume the available output tensor is structured and complete. However, in many applications, these assumptions are not valid, and the measurements may contain missing values for reasons such as the failure of the sensors or excessive cost of full measurement. For example, in lithography process of semiconductor manufacturing, overlay (OV) errors (i.e., the misalignment between different layers), which are highly dependent on the lithography machine settings (e.g., the alignments of the lens and the location of the wafer stage), are only measured at a limited number of marked locations over the wafer (see Fig. 1). Therefore, obtaining a full picture of OV errors requires developing models that link the overlays to machine settings based on partially observed OV errors on wafers.

Another example is the battery system of the Tesla Model S, with more than 7000 cells [15], where a limited number of sensors monitors the temperature of a few of these cells. As a result, the temperature data obtained from these batteries is structured but contains missing observations. To effectively monitor the battery condition, a modeling framework that uses this data to estimate the temperature of all cells based on the observed measurements and covariates such as the speed and temperature of coolant flow in the battery pack is essential.

In these situations, where the output contains missing values, an effective estimation of the model parameters is challenging due to the absence of several observations in high-dimensional outputs, which renders the exploitation of the complex correlation structure of the output even more difficult. A naïve approach of addressing this challenge is to first complete the HD outputs, using matrix [16] or tensor completion approaches [17], and then construct a prediction model based on the completed data using existing HD regression methods (e.g., [14]). Many methods are developed for the tensor completion problem, which attempt to build the relationship between the known entries and the missing values of a matrix or a tensor. These methods rely on the low-rank assumption of the tensor and either design decomposition of the tensor [18], [19], or perform rank minimization [17], [20], as reported in Table II. Nevertheless, these approaches do not consider other potentially relevant and available data when performing the tensor completion task. In many applications, however, the incomplete tensor is related to other variables (e.g., the inputs of a process) that can provide further information in the completion procedure.

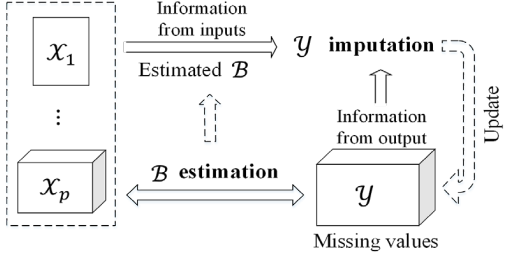

Fig. 2. Proposed framework of the augmented regression modeling with an incomplete response.

To fully exploit the available information, we propose an augmented tensor regression framework that simultaneously estimates the model parameters and completes the response tensor. Fig. 2 illustrates an overview of our procedure. In this figure, $\mathcal{X}_j, j = 1, \cdots, p$ represent the available inputs that are complete, and $\mathcal{Y}$ denotes an incomplete output of a process. Please note that in the situations where $\mathcal{X}_j, j = 1, \cdots, p$ contain missing values, one can complete them using tensor completion approaches as they are independent variables. The goal is to estimate the model parameters $\mathcal{B}$, given the inputs and incomplete output. After learning the model parameters, a complete output can be estimated based on a given set of new inputs. As it is shown in Fig. 2, our procedure completes the output and learns the model parameters iteratively. Specifically, the completion of the response $\mathcal{Y}$ takes advantage of both the information of the inputs via the estimated parameters $\mathcal{B}$ and the structural information in the response. By considering the information from inputs, the completion of response is expected to be more accurate than a completion procedure that only uses the structural information of $\mathcal{Y}$. Furthermore, a better completion of the response $\mathcal{Y}$ can also benefit the estimation of the model parameters. Therefore, all available information from both inputs and output are fully and iteratively exploited, resulting in more accurate model estimation and improved performance of prediction.

Mathematically, the model parameters are learned by minimizing an integrated objective function that includes a mean square error term and a tensor rank penalty on the response $\mathcal{Y}$ for global exploitation of the relationship between the observed and missing entries. To avoid overfitting, a decomposition operation is applied on the model parameters $\mathcal{B}$ by considering the correlation structures within the inputs and output spaces. The optimization problem is then solved through a novel and efficient algorithm with two block coordinate descent (BCD) steps. The proposed optimization framework iteratively completes the output tensor and learns the model parameters until convergence.

The rest of the article is organized as follows: In Section II, we propose an integrated framework for augmented tensor regression with missing values and elaborate on the algorithms for estimating the missing values of the response as well as the model parameters. Section III carries out two simulation studies. The first one implements a curve-on-curve simulation with missing entries in the profile output. The second simulation study generates images and profiles for the inputs and the output. Based on those two simulations, the proposed method is evaluated in comparison to a benchmark method. In Section IV, we conduct a case study to estimate the incomplete overlay errors in the semiconductor lithographic process. Finally, we conclude the paper in Section V.

## II. FORMULATION OF TENSOR REGRESSION WITH INCOMPLETE RESPONSE

In this section, we propose an approach that augments tensor-on-tensor regression with tensor completion for a more accurate estimation of model parameters and the high-dimensional process output with missing values. First, we introduce the notations and concepts of tensor algebra used in this paper.

### A. Tensor Notation and Multilinear Algebra

In this paper, we use Euler script letters to denote a tensor. For instance, $\mathcal{X} \in \mathbb{R}^{P_1 \times P_2 \times \cdots \times P_n}$ denotes a tensor of order $n$, where $P_i$ indicates the dimension of the mode $i$ of the tensor $\mathcal{X}$. The *mode-$i$ matricization* and the *vectorization* of a tensor $\mathcal{X}$ are denoted by $\mathbf{X}_{(i)} \in \mathbb{R}^{P_i \times P_{-i}}$ ( $P_{-i} = P_1 \times P_2 \times \cdots \times P_{i-1} \times P_{i+1} \times \cdots \times P_n$) and $\text{vec}(\mathcal{X})$, respectively. A tensor $\mathcal{X}$ can be obtained by *folding* any of its matricizations, which is denoted as $\mathcal{X} = \text{fold}(\mathbf{X}_{(i)}), i \in \{1, \ldots, n\}$. The *mode-$k$ product* of a tensor $\mathcal{X}$ with a matrix $\mathbf{U} \in \mathbb{R}^{K \times P_k}$ is denoted by $\mathcal{X} \times_k \mathbf{U} \in \mathbb{R}^{P_1 \times P_2 \times \cdots \times P_{k-1} \times K \times P_{k+1} \times \cdots \times P_n}$, where the $p_1 \cdots p_{k-1} k p_{k+1} \cdots p_n$ entry of the product is given by,

$$(\mathcal{X} \times_k \mathbf{U})_{p_1 \cdots p_{k-1} k p_{k+1} \cdots p_n} = \sum_{p_k=1}^{P_k} x_{p_1 \cdots p_n} u_{j p_k}.$$

The *Frobenius norm* of a tensor $\mathcal{X}$ is defined as the *Frobenius norm* of its matricization along any mode, e.g., $\|\mathcal{X}\|_F^2 = \|\mathbf{X}_{(i)}\|_F^2$. The *contraction product* of two tensors $\mathcal{B} \in \mathbb{R}^{P_1 \times P_2 \times \cdots \times P_n \times Q_1 \times \cdots \times Q_d}$ and $\mathcal{X} \in \mathbb{R}^{P_1 \times P_2 \times \cdots \times P_n}$ is denoted by $\mathcal{B} * \mathcal{X} \in \mathbb{R}^{Q_1 \times \cdots \times Q_d}$,

$$(\mathcal{B} * \mathcal{X})_{q_1 \cdots q_d} = \sum_{p_1 \cdots p_l} \mathcal{X}_{p_1 \cdots p_l} \mathcal{B}_{p_1 \cdots p_l q_1 \cdots q_d}.$$

We use $\|\mathbf{X}\|_* = \sum_i \lambda_i(\mathbf{X})$ to denote the *nuclear norm* of the matrix $\mathbf{X}$, where $\lambda_i(\mathbf{X})$ is the $i^{th}$ largest singular value of the matrix. The nuclear norm of a tensor $\mathcal{X}$, denoted by $\|\mathcal{X}\|_*$, is defined as the weighted average of nuclear norms of its matricizations along each mode [17], i.e., $\|\mathcal{X}\|_* = \sum_{i=1}^d \alpha_i \|\mathbf{X}_{(i)}\|_*$, where $\alpha_i > 0$ and $\sum_i \alpha_i = 1$. We let $\Omega$ denote an index set, and $\mathcal{P}_\Omega(\mathcal{X})$ represent an orthogonal projection that copies the tensor entries of $\mathcal{X}$ with indices in $\Omega$ and sets the entries with indices outside $\Omega$ to be 0. The *Tucker rank* of tensor $\mathcal{X}$ is defined as the combination of the column ranks of its matricizations. That is, $\text{rank}(\mathcal{X}) = (R_1, R_2, \cdots, R_n)$, where $R_i$ is the column rank of $\mathbf{X}_{(i)}$ [21].

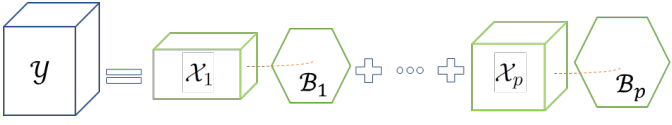

Fig. 3. Illustration of the regression model for heterogeneous sets of data.

### B. Problem Formulation

Let $m$ denote the number of available samples in the training set, $\mathcal{Y}_{0i} \in \mathbb{R}^{Q_1 \times Q_2 \times \cdots \times Q_d}$ $(i = 1, \cdots, m)$ be the response tensor in the $i^{th}$ sample that contains missing values, and $\mathcal{X}_{ji} \in \mathbb{R}^{P_{j1} \times P_{j2} \times \cdots \times P_{jl_j}}$ $(i = 1, \cdots, m; j = 1, \cdots, p)$ represents the $j^{th}$ input tensor (predictor) within the $i^{th}$ sample. Let us denote by $\Omega_i$ the set of indices at which entries of $\mathcal{Y}_{0i}$ are observed. Then, we characterize the relationship between the inputs and the output by,

$$\mathcal{P}_{\Omega_i}(\mathcal{Y}_{0i}) = \mathcal{P}_{\Omega_i}\left(\sum_{j=1}^{p} \mathcal{X}_{ji} * \mathcal{B}_j\right) + \mathcal{P}_{\Omega_i}(\mathcal{E}_i), \tag{1}$$
$$i = 1, \cdots, m,$$

where $\mathcal{B}_j \in \mathbb{R}^{P_{j1} \times P_{j2} \times \cdots \times P_{jl_j} \times Q_1 \times Q_2 \times \cdots \times Q_d}$ is the tensor of coefficients related to input $j$ and $\mathcal{E}_i$ is a tensor of errors whose entries follow a random process. For a more compact representation, we can fold tensors $\mathcal{Y}_i$, $\mathcal{X}_{ji}$, and $\mathcal{E}_i$ $(i = 1, \cdots, m)$ along the sample mode to obtain tensors $\mathcal{Y}_0 \in \mathbb{R}^{m \times Q_1 \times Q_2 \times \cdots \times Q_d}$, $\mathcal{X}_j \in \mathbb{R}^{m \times P_{j1} \times P_{j2} \times \cdots \times P_{jl_j}} (j = 1, \cdots, p)$, and $\mathcal{E} \in \mathbb{R}^{m \times Q_1 \times Q_2 \times \cdots \times Q_d}$. Then, the model can be written as,

$$\mathcal{P}_\Omega(\mathcal{Y}_0) = \mathcal{P}_\Omega\left(\sum_{j=1}^{p} \mathcal{X}_j * \mathcal{B}_j\right) + \mathcal{P}_\Omega(\mathcal{E}), \tag{2}$$

where $\Omega = \{(i, q_1, \cdots, q_d) \in \mathbb{R}^{m \times Q_1 \times Q_2 \times \cdots \times Q_d}\}$ is a set of indices at which entries of $\mathcal{Y}_0$ are observed. Upon estimation of $\mathcal{B}_j$, the complete output can be estimated by $\sum_{j=1}^{p} \mathcal{X}_j * \mathcal{B}_j$, as illustrated in Fig. 3. Because the output contains missing values, existing approaches that assume complete tensors are not applicable for estimating the parameters. Fortunately, due to the potential correlation structure within high-dimensional data points, including profiles and images, the output tensor generally has a low-rank structure. This low-rank structure is key in designing an approach for estimating the model parameters in the presence of incomplete output tensor. The goal of our work is to build a regression model between the input tensors $\mathcal{X}_j, (j = 1, \cdots, p)$ and the incomplete output tensor $\mathcal{Y}_0$. It can be achieved by integrating the low-rank structure of the response into the least square loss $L_r$ as follows,

$$\min_{\mathcal{Y}, \mathcal{B}_1, \cdots, \mathcal{B}_p} \lambda \operatorname{rank}(\mathcal{Y}) + \frac{1}{2} \left\| \mathcal{Y} - \sum_{j=1}^{p} \mathcal{X}_j * \mathcal{B}_j \right\|_{\mathrm{F}}^2 \tag{3}$$

$$\text{Subject to: } \mathcal{P}_\Omega(\mathcal{Y}) = \mathcal{P}_\Omega(\mathcal{Y}_0).$$

The first term, $\lambda \operatorname{rank}(\mathcal{Y})$, in the objective function, exploits the low-rank structure of the output tensor. Note that the assumption of this problem is that the output is low-rank, which is common in high dimensional spaces and can be validated based on the domain knowledge. $\lambda$ is a user-specified tuning parameter that should be selected based on the procedure discussed in Section II.E. Unfortunately, problem (3) is NP-hard due to the nonconvexity of $\operatorname{rank}(\mathcal{Y})$. To address this issue, we employ a convex relaxation of the $\operatorname{rank}(\mathcal{Y})$ term. Specifically, we employ the tensor's nuclear norm to approximate the rank penalty, leading to the following tractable problem,

$$\min_{\mathcal{Y}, \mathcal{B}_1, \cdots, \mathcal{B}_p} \lambda \|\mathcal{Y}\|_* + \frac{1}{2} \left\| \mathcal{Y} - \sum_{j=1}^{p} \mathcal{X}_j * \mathcal{B}_j \right\|_{\mathrm{F}}^2 \tag{4}$$

$$\text{Subject to: } \mathcal{P}_\Omega(\mathcal{Y}) = \mathcal{P}_\Omega(\mathcal{Y}_0),$$

where $\|\mathcal{Y}\|_* = \sum_{i=1}^{d} \alpha_i \|\mathbf{Y}_{(i)}\|_*$ as it is defined in Section II.A. Solving this problem without imposing any constraint over the parameters $\mathcal{B}_j$ will result in severe overfitting as the number of parameters are extremely large. To avoid this issue, we decompose $\mathcal{B}_j, (j = 1, \cdots, p)$ by using a Tucker operation and write,

$$\mathcal{B}_j = \mathcal{C}_j \times_1 \mathbf{U}_{j1} \times_2 \cdots \times_{l_j} \mathbf{U}_{jl_j} \times_{l_j+1} \mathbf{V}_1 \times_{l_j+2} \cdots \times_{l_j+d} \mathbf{V}_d.$$

Here, $\mathcal{C}_j \in \mathbb{R}^{\tilde{P}_{j1} \times \cdots \times \tilde{P}_{jl_j} \times \tilde{Q}_1 \times \cdots \times \tilde{Q}_d}$ is a core tensor with $\tilde{P}_{ji} \ll P_{ji}$ $(j = 1, \cdots, p; i = 1, \cdots, l_j)$ and $\tilde{Q}_i \ll Q_i$ $(i = 1, \cdots, d)$; $\{\mathbf{U}_{ji} : j = 1, \cdots, p; i = 1, \cdots, l_j\}$ is a set of bases that spans the $j^{th}$ input space; and $\{\mathbf{V}_i : i = 1, \cdots, d\}$ is a set of bases that spans the output space. Therefore, we aim to solve,

$$\min_{\mathcal{Y}, \mathcal{B}_1, \cdots, \mathcal{B}_p} \lambda \|\mathcal{Y}\|_* + \frac{1}{2} \left\| \mathcal{Y} - \sum_{j=1}^{p} \mathcal{X}_j * \mathcal{B}_j \right\|_{\mathrm{F}}^2,$$

$$\text{Subject to: } \mathcal{P}_\Omega(\mathcal{Y}) = \mathcal{P}_\Omega(\mathcal{Y}_0); \quad \mathcal{B}_j = \tag{5}$$

$$\mathcal{C}_j \times_1 \mathbf{U}_{j1} \times_2 \ldots \times_{l_j} \mathbf{U}_{jl_j} \times_{l_j+1} \mathbf{V}_1 \times_{l_j+2} \cdots \times_{l_j+d} \mathbf{V}_d,$$
$$j = 1, \ldots, p.$$

To estimate the model parameters, we design an algorithm with the following two steps (see **Algorithm 1**) as follows:

Step 1 ($\mathcal{Y}$-update): This step optimizes the following sub-problem assuming that the model parameters $\mathcal{B}_j^k$ $(j = 1, \cdots, p)$ are given at the $(k + 1)^{th}$ iteration of the algorithm,

$$\mathcal{Y}^{k+1} = \operatorname{argmin}_{\mathcal{Y}} \lambda \|\mathcal{Y}\|_* + \frac{1}{2} \left\| \mathcal{Y} - \sum_{j=1}^{p} \mathcal{X}_j * \mathcal{B}_j^k \right\|_{\mathrm{F}}^2 \tag{6}$$

$$\text{Subject to: } \mathcal{P}_\Omega(\mathcal{Y}) = \mathcal{P}_\Omega(\mathcal{Y}_0).$$

Let $\mathcal{A}^k = \sum_{j=1}^{p} \mathcal{X}_j * \mathcal{B}_j^k$; then the problem can be rewritten as,

---

**Algorithm 1:** BCD algorithm for solving problem (5)

---

1: **Inputs:** $\mathcal{X}_1, \ldots, \mathcal{X}_p, \mathcal{Y}_0$ and $\Omega$.
   Initiate $\mathcal{B}_1^0, \ldots, \mathcal{B}_p^0$ randomly and let $\mathcal{Y}^0 = \mathcal{Y}_0$.
2: **Loop**
3:    $\mathcal{Y}^{k+1}$ *$\mathcal{Y}$-update* step: update $\mathcal{Y}^k$ to $\mathcal{Y}^{k+1}$ by solving
      Equation (7) given $\mathcal{B}_j^k$ ($j = 1, \cdots, p$)
4:    Let $\mathcal{Y}^k = \mathcal{Y}^{k+1}$
5:    **for** $j = 1, \cdots, p$ **do**:
6:      $\mathcal{B}_j^{k+1}$ *$\mathcal{B}_j$-update* step: update $\mathcal{B}_j^k$ to $\mathcal{B}_j^{k+1}$ by solving
        Equation (8) given $\mathcal{Y}^k$ and the remaining $\mathcal{B}_i^k$ for
        $i = 1, \cdots, p, \ i \neq j$.
7:    Let $\mathcal{B}_j^k = \mathcal{B}_j^{k+1}$
8    **end**
9: **until** convergence.

---

$$\mathcal{Y}^{k+1} = \arg\min_{\mathcal{Y}} \lambda \|\mathcal{Y}\|_* + \frac{1}{2} \|\mathcal{Y} - \mathcal{A}^k\|_F^2 \tag{7}$$

$$\text{Subject to: } \mathcal{P}_\Omega(\mathcal{Y}) = \mathcal{P}_\Omega(\mathcal{Y}_0).$$

<u>Step 2 ($\mathcal{B}_j$-update)</u>: Given the estimated $\mathcal{Y}^k$ and remaining $\mathcal{B}_i^k$ ($i = 1, \cdots, p, \ i \neq j$), this step attempts to solve,

$$\mathcal{B}_j^{k+1} = \arg\min_{\mathcal{B}_j} \|\mathcal{W}_j^k - \mathcal{X}_j * \mathcal{B}_j\|_F^2$$

Subject to: $\mathcal{B}_j = $ \hfill (8)

$$\mathcal{C}_j \times_1 \mathbf{U}_{j1} \times_2 \cdots \times_{l_j} \mathbf{U}_{jl_j} \times_{l_j+1} \mathbf{V}_1 \times_{l_j+2} \cdots \times_{l_j+d} \mathbf{V}_d,$$

where $\mathcal{W}_j^k = \mathcal{Y}^k - \sum_{i \neq j}^p \mathcal{X}_i * \mathcal{B}_i^k$.

The stopping criterion of **Algorithm 1** is defined by the maximum number of iterations.

### C. Solution to $\mathcal{Y}$-update

The problem of $\mathcal{Y}$-update (i.e., Equation (7)) can be transformed in,

$$\min_{\mathbf{Y}_{(1)}, \cdots, \mathbf{Y}_{(d)}} \sum_{i=1}^d \alpha_i \left\{ \lambda \|\mathbf{Y}_{(i)}\|_* + \frac{1}{2} \|\mathbf{Y}_{(i)} - \mathbf{A}_{(i)}\|_F^2 \right\} \tag{9}$$

$$\text{Subject to: } \mathcal{P}_\Omega(\mathcal{Y}) = \mathcal{P}_\Omega(\mathcal{Y}_0)$$

The details of this transformation are provided in APPENDIX A. The default values of parameter $\alpha_i, i = 1, \ldots, d$ are set to $\frac{1}{d}$. That is, we assign equal weights to each of the tensor modes in the nuclear norm.

Since this problem is the same at each iteration, we use the general notation $\mathcal{Y}$ and $\mathcal{A}$ by ignoring the iteration index $k$ for simplicity. Solving the above problem is challenging due to the interdependent nuclear norm terms. That is, the entries of the matricizations of the output tensor in the nuclear norm are shared and therefore, cannot be treated independently while minimizing the objective function. To tackle this issue, we define several auxiliary variables to split these interdependent terms. Specifically, we introduce additional local matrices $\mathbf{M}_{(1)}, \cdots, \mathbf{M}_{(d)}$, which represent the mode-$i$ matricizations of local tensors $\mathcal{M}_1, \cdots, \mathcal{M}_d, i = 1, 2, \ldots, d$, separately. Then, we have the following problem,

$$\min_{\mathbf{M}_{(1)}, \cdots, \mathbf{M}_{(d)}, \mathcal{Y}} \sum_{i=1}^d \alpha_i \left\{ \lambda \|\mathbf{M}_{(i)}\|_* + \frac{1}{2} \|\mathbf{M}_{(i)} - \mathbf{A}_{(i)}\|_F^2 \right\}$$

$$\text{Subject to: } \mathcal{P}_\Omega(\mathcal{Y}) = \mathcal{P}_\Omega(\mathcal{Y}_0) \tag{10}$$

$$\mathcal{Y} = \mathcal{M}_i, i = 1, \cdots, d.$$

Here, $\mathcal{Y}$ can be treated as a global variable, and the problem becomes a global consensus problem. By introducing $d$ local variables, problem (10) becomes separable since they are independent of each other. To solve (10), we first address the equality constraints $\mathbf{M}_{(i)} = \mathbf{Y}_{(i)}$ ($i = 1, \cdots, d$) by defining the augmented Lagrangian as follows,

$$L_\rho(\mathbf{M}_{(1)}, \cdots, \mathbf{M}_{(d)}, \mathcal{Y}, \Theta_1, \cdots, \Theta_d)$$

$$= \sum_{i=1}^d \begin{Bmatrix} \lambda \alpha_i \|\mathbf{M}_{(i)}\|_* + \frac{\alpha_i}{2} \|\mathbf{M}_{(i)} - A_{(i)}\|_F^2 + \\ \langle \Theta_i, \mathcal{Y} - \mathcal{M}_i \rangle + \frac{\rho}{2} \|\mathcal{Y} - \mathcal{M}_i\|_F^2 \end{Bmatrix}, \tag{11}$$

where $\langle \Theta_i, \mathcal{Y} - \mathcal{M}_i \rangle$ represents inner product of tensors $\Theta_i$ and $\mathcal{Y} - \mathcal{M}_i$; $\Theta_i$ ($i = 1, \ldots, d$) denote the tensors of dual variables; and $\rho$ is the step size. Then, the problem is represented as,

$$\min_{\mathbf{M}_{(1)}, \cdots, \mathbf{M}_{(d)}, \mathcal{Y}, \Theta_1, \cdots, \Theta_d} L_\rho(\mathbf{M}_{(1)}, \cdots, \mathbf{M}_{(d)}, \mathcal{Y}, \Theta_1, \cdots, \Theta_d)$$

$$\text{Subject to: } \mathcal{P}_\Omega(\mathcal{Y}) = \mathcal{P}_\Omega(\mathcal{Y}_0). \tag{12}$$

The resulting ADMM algorithm for this problem is summarized in **Algorithm 2**, which includes three iterative steps: local variable update, global variable update, and dual variable update. For local variable update, we try to solve the following unconstrained problem,

$$\min_{\mathbf{M}_{(i)}} f_i(\mathbf{M}_{(i)}), \tag{13}$$

where $f_i(\mathbf{M}_{(i)}) = \lambda \alpha_i \|\mathbf{M}_{(i)}\|_* + \frac{\alpha_i}{2} \|\mathbf{M}_{(i)} - \mathbf{A}_{(i)}\|_F^2 + \langle \Theta_i, \mathcal{Y} - \mathcal{M}_i \rangle + \frac{\rho}{2} \|\mathcal{Y} - \mathcal{M}_i\|_F^2$.

This problem is equivalent to,

$$\min_{\mathbf{M}_{(i)}} \lambda_i \|\mathbf{M}_{(i)}\|_* + \frac{1}{2} \|\mathbf{M}_{(i)} - \mathbf{C}_{(i)}\|_F^2, \tag{14}$$

where $\lambda_i = \frac{\lambda \alpha_i}{\alpha_i + \rho}$ and $\mathbf{C}_{(i)} = \frac{\alpha_i \mathbf{A}_{(i)} + \rho \mathbf{Y}_{(i)} + \Theta_{(i)}}{\alpha_i + \rho}$.

The detailed derivation of this problem is provided in APPENDIX B. It turns out that this minimization problem in Equation (14) can be solved by first computing the singular value decomposition (SVD) of $\mathbf{C}_{(i)}$ and then applying a soft-thresholding operator on the singular values. This result is proved by [16] and presented in the following proposition.

**Proposition 1**. Let $\mathbf{C} \in \mathbb{R}^{m \times n}$ and let $\mathbf{C} = \mathbf{U}\boldsymbol{\Sigma}\mathbf{V}^T$ be the SVD of $\mathbf{C}$, where $\mathbf{U} \in \mathbb{R}^{m \times m}$ and $\mathbf{V} = \mathbb{R}^{n \times n}$ are orthonormal matrices, $\boldsymbol{\Sigma} \in \mathbb{R}^{m \times n}$ is a diagonal matrix, and $r = \mathrm{rank}(\mathbf{C})$. Then,

$$\widehat{\mathbf{M}} \equiv \underset{\mathbf{M}}{\mathrm{argmin}} \left\{ \lambda \|\mathbf{M}\|_* + \frac{1}{2}\|\mathbf{M} - \mathbf{C}\|_F^2 \right\}$$

is optimized by $\widehat{\mathbf{M}} = \mathbf{U}_r \boldsymbol{\Sigma}_\lambda \mathbf{V}_r^T$, where $\boldsymbol{\Sigma}_\lambda$ is diagonal with $(\boldsymbol{\Sigma}_\lambda)_{ii} = \max(0, \boldsymbol{\Sigma}_{ii} - \lambda)$, $\mathbf{U}_r$ and $\mathbf{V}_r$ are the first $r$ columns of $\mathbf{U}$ and $\mathbf{V}$.

In order to update the global variable, we attempt to solve the following constrained optimization problem,

$$\min_{\mathcal{Y}} g(\mathcal{Y}) \tag{15}$$

$$\text{Subject to: } \mathcal{P}_\Omega(\mathcal{Y}) = \mathcal{P}_\Omega(\mathcal{Y}_0),$$

where $g(\mathcal{Y}) = \sum_{i=1}^d \left\{ \langle \Theta_i, \mathcal{Y} - \mathcal{M}_i \rangle + \frac{\rho}{2}\|\mathcal{Y} - \mathcal{M}_i\|_F^2 \right\}$.

First, we compute its Lagrangian as follows,

$$L_y(\mathcal{Y}, \Phi) = \sum_{i=1}^d \left\{ \langle \Theta_i, \mathcal{Y} - \mathcal{M}_i \rangle + \frac{\rho}{2}\|\mathcal{Y} - \mathcal{M}_i\|_F^2 \right\} \\ + \langle \Phi, \mathcal{P}_\Omega(\mathcal{Y} - \mathcal{Y}_0) \rangle, \tag{16}$$

where $\Phi$ is the tensor of the dual variable in this sub-problem.

Then, a closed-form solution for this sub-problem is derived based on the optimality condition,

$$\mathcal{Y} = \begin{cases} \mathcal{Y}_0, & \text{if } (j, q_1, \cdots, q_d) \in \Omega \\ \frac{1}{d}\sum_{i=1}^d \left( \mathcal{M}_i - \frac{1}{\rho}\Theta_i \right), & \text{otherwise} \end{cases} \tag{17}$$

The detailed derivations are provided in APPENDIX C. Finally, the dual variables $\Theta_i$ are updated separately according to,

$$\Theta_i \leftarrow \Theta_i + \rho(\mathcal{Y} - \mathcal{M}_i). \tag{18}$$

### D. Solution to $\mathcal{B}$-update

After the estimation of the response $\mathcal{Y}$ in each iteration of **Algorithm 1**, we then estimate the tensor of coefficients $\mathcal{B}_j$. For this problem, the key is to learn the core tensor $\mathcal{C}_j$ and the bases $\mathbf{V}_i$ and $\mathbf{U}_{ji}$. Motivated by [14], we learn $\mathbf{U}_{ji}$ directly from the input spaces by performing Tucker decomposition on the input tensors. After the estimation of $\mathbf{U}_{ji}$, the problem of learning $\mathcal{B}_j$ becomes the following optimization problem,

$$\{\mathcal{C}_j, \mathbf{V}_1, \cdots, \mathbf{V}_d\} = \underset{\mathcal{B}_j}{\mathrm{argmin}}\|\mathbf{W}_{j(1)} - \mathbf{X}_{(1)}\mathbf{B}_j\|_F^2$$

$$\text{Subject to: } \mathcal{B}_j =$$

$$\mathcal{C}_j \times_1 \mathbf{U}_{j1} \times_2 \cdots \times_{l_j} \mathbf{U}_{jl_j} \times_{l_j+1} \mathbf{V}_1 \times_{l_j+2} \cdots \times_{l_j+d} \mathbf{V}_d,$$

$$\mathbf{V}_i^T\mathbf{V}_i = \mathbf{I}_{\tilde{Q}_i}(i = 1, \cdots, d), \tag{19}$$

where $\mathbf{W}_{j(1)}$ and $\mathbf{X}_{(1)}$ are mode-1 matricizations of $\mathcal{W}_j$ and $\mathcal{X}_j$,

---

**Algorithm 2** ADMM algorithm for problem (12)

Inputs: $\mathcal{X}_1, \ldots, \mathcal{X}_p$, the estimated $\widehat{\mathcal{B}}_p^{k-1}, \ldots, \widehat{\mathcal{B}}_p^{k-1}, \mathcal{Y}_0$ and $\Omega$.
Loop:
  For $i = 1, \ldots, d$:
    $\mathbf{M}_{(i)}^k \leftarrow \arg\min_{\mathbf{M}_{(i)}} f_i(\mathbf{M}_{(i)})$

    $\mathcal{M}_i^k \leftarrow \mathrm{fold}(\mathbf{M}_{(i)}^k)$       (Local variable)

    $\mathcal{Y}^k \leftarrow \arg\min_{\{\mathcal{Y}:\, \mathcal{P}_\Omega(\mathcal{Y}) = \mathcal{P}_\Omega(\mathcal{Y}_0)\}} g(\mathcal{Y})$    (Global variable)

  For $i = 1, \ldots, d$:
    $\Theta_i^k \leftarrow \Theta_i^{k-1} + \rho(\mathcal{Y}^k - \mathcal{M}_i^k)$    (Dual variable)

Until convergence.

---

respectively. $\mathbf{B}_j \in \mathbb{R}^{P_j \times Q}$ is the matricization of tensor $\mathcal{B}_j$ with $P_j = \prod_{k=1}^{l_j} P_{jk}$ and $Q = \prod_{k=1}^d Q_k$. $\mathbf{I}_{\tilde{Q}_i} \in \mathbb{R}^{\tilde{Q}_i \times \tilde{Q}_i}$ is an identity matrix. The problem (19) can be solved by using the ALS-BCD as in [14].

One requirement of ALS-BCD algorithm is to know the rank of the response $\mathcal{Y}$. Since $\mathcal{Y}$ is estimated during the $\mathcal{Y}$-*update* step in **Algorithm 1**, we use high-order SVD (HOSVD) to estimate its rank, which is the column rank of $\mathbf{Y}_{(n)}$ for $n = 1, \cdots, d$. More specifically, we use the truncated SVD to estimate the rank of mode-n matricization $\mathbf{Y}_{(n)}$ by specifying a ratio, by which the data variability is explained. Then, the ranks of the matricizations of $\mathcal{Y}$ along all modes are estimated, which are then combined into the Tucker rank of tensor $\mathcal{Y}$ for ALS-BCD algorithm.

### E. Selection of Hyper-parameters

In the proposed method, the global hyper-parameter $\lambda$ and local hyper-parameters $\tilde{P}_{ji}$ $(j = 1,2,\cdots,p; i = 1,2,\cdots,l_j)$ should be identified. $\lambda$ is designed to balance the rank penalty and the least square error in the objective function. Based on our experiments, the proposed method is robust to a wide range of values of $\lambda$. In this paper, we set $\lambda = 1$ by default. This parameter can also be determined by cross-validation.

Similar to the procedure for estimating the rank of $\mathcal{Y}$ in Section II.D, the Tucker rank of input tensors $\mathcal{X}_j$ can also be estimated using the truncated HOSVD. Based on the estimated rank of input tensors, the parameters $\tilde{P}_{ji}$ $(j = 1,2,\cdots,p; i = 1,2,\cdots,l_j)$ are determined.

## III. SIMULATION STUDY

In this section, a set of simulation studies is carried out to evaluate the performance of our proposed method, herein referred to as the tensor regression with missing values (called TRMV). A benchmark method, named TC-MTOT, is used for comparison, where the completion of output tensor (TC) is first conducted with the rank minimization method [16], [17] and then the regression model between the inputs and the completed response is constructed using MTOT [14]. The estimated models by TRMV and TC-MTOT are used to predict an output given a new set of inputs. Finally, the prediction error is calculated for quantitative comparison.

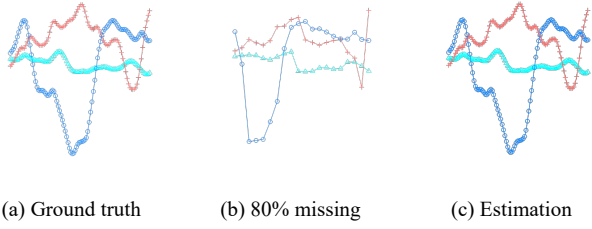

(a) Ground truth    (b) 80% missing    (c) Estimation

Fig. 4. Examples of the output profiles, including the original profiles (100 observations), the profiles with 80% missing values (20 observations), and the estimated ones.

### A. Curve-on-curve Regression

Profile-on-profile regression has a wide range of applications in both manufacturing and healthcare [22]. This simulation mimics profile-on-profile regression modeling when the output contains missing values. Following the simulation study in [23], we first randomly generate $\left(\mathcal{B}_1, \cdots, \mathcal{B}_p\right)$ as follows:

$$\mathcal{B}_i(s,t) = \frac{1}{p^2}\left[\gamma_{1i}(t)\psi_{1i}(s) + \gamma_{2i}(t)\psi_{2i}(s) + \gamma_{3i}(t)\psi_{3i}(s)\right],$$

where $\gamma_{ki}(t)$ and $\psi_{ki}(s)$ $(k=1,2,3; i=1,\cdots,p)$ are Gaussian processes with covariance function $\Sigma_1(z,z') = \left(1 + 20|z-z'| + \frac{1}{3}(20|z-z'|)^2\right)e^{-20|z-z'|}$. Then, $p$ input profiles are simulated with the following three steps: 1) generate a matrix $\mathbf{S} \in \mathbb{R}^{p \times p}$ with the $(i,j)^{th}$ entry equal to 1 for $i = j$ and equal to $\rho_c = 0$ or $0.5$ for $i \neq j$ ; 2) apply eigendecomposition $\mathbf{S} = \Delta\Delta^{\mathrm{T}}$ to obtain a $p \times p$ matrix $\Delta$ and generate a set of profiles $\mathbf{w}_1, \mathbf{w}_2, \cdots, \mathbf{w}_p$ by using a Gaussian process with covariance function $\Sigma_2(z,z') = e^{-2|z-z'|^2}$ ; 3) generate the input curves at any given point $s$ by,

$$\left(\mathbf{x}_1(s), \cdots, \mathbf{x}_p(s)\right) = \left(\mathbf{w}_1(s), \cdots, \mathbf{w}_p(s)\right)\Delta^{\mathrm{T}}.$$

According to this procedure, for each $s$ the vector $\left(x_1(s), \cdots, x_p(s)\right)$ follows a multivariate Gaussian distribution with covariance $\mathbf{S}$. When setting $\rho_c = 0$ to generate data, the variables of the vector $\left(x_1(s), \cdots, x_p(s)\right)$ are uncorrelated. The output profiles can finally be simulated by,

$$y(t) = \sum_{j=1}^{p} \int \mathcal{B}_j(s,t)x_j(s)ds + \epsilon(t),$$

where $\epsilon(t)$ are independently distributed random variables from a Gaussian distribution with mean of zero and the variance of $\sigma^2$. All the input and output profiles are generated over an equidistant grid of size 100, which are defined over the intervals $0 < s < 2$ and $0 < t < 1$, respectively. In this study, we set $p = 2$. After the profiles are generated, we randomly remove $r = 80\%$ and 90% of the entries of the output profile in the training dataset. We refer to this data generation steps as *procedure A*. Fig. 4 shows an example of the ground truth

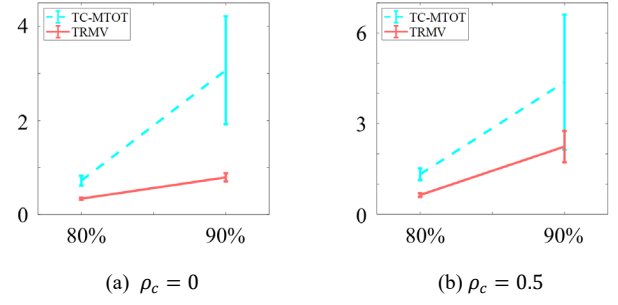

(a) $\rho_c = 0$      (b) $\rho_c = 0.5$

Fig. 5. Performance comparison between our proposed method (TRMV) and TC-MTOT under different levels of missing values.

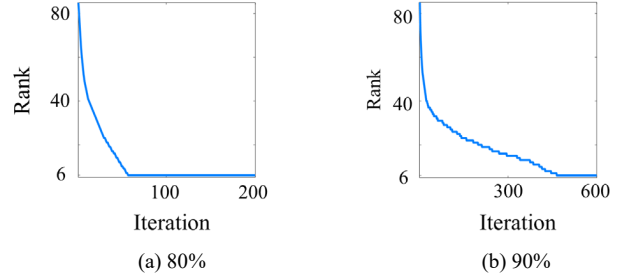

(a) 80%      (b) 90%

Fig. 6. Process of rank estimation for the response at different levels of missing values.

profile, the profile with 80% missing values, and its estimation by TRMV.

Based on *procedure A*, we first generate a dataset of 200 samples with $\sigma = 0$. We use $m_{tr} = 100$ samples for model training and hyper-parameter estimations and the remaining $m_{ts} = 100$ samples for testing. For quantitative evaluation, the standardized prediction error (SPE) is used to evaluate the performance of the proposed model and is defined as $SPE = \frac{\|y - \hat{y}\|_F}{\|y\|_F}$. For better illustration, we transform SPE by taking the negative inverse of its logarithm, i.e., $-\frac{1}{log(SPE)}$, called transformed SPE (TSPE).

Given the generated dataset, we employ TRMV and TC-MTOT methods to complete $y$ and estimate the tensor of coefficients $\mathcal{B}_j$ $(j = 1,2)$ at the aforementioned levels of missing values. The trained models are then applied to the testing set to predict responses. The performance of each of the models is then evaluated and compared in terms of TSPE. Fig. 5 compares the performance of TRMV to TC-MTOT at different levels of missing values in the response. The results reveal that TRMV outperforms TC-MTOT at every level of missing values (LMV). For example, at 80% missing values for $\rho_c = 0$, the mean TSPE of the proposed method is 0.3382, which is significantly lower than that of TC-MTOT, i.e., 0.7206. This indicates that TRMV can to take advantage of the available information in both the inputs and the response, leading to improved estimations.

Fig. 6 shows the process of rank estimation. As it is illustrated, when the proportion of missing values is relatively small, our algorithm converges to the true rank with fewer iterations. For example, less than 60 iterations are sufficient to find the true rank for 80% missing values. In contrast, it takes about 480 iterations to converge to the true rank for 90% missing values.

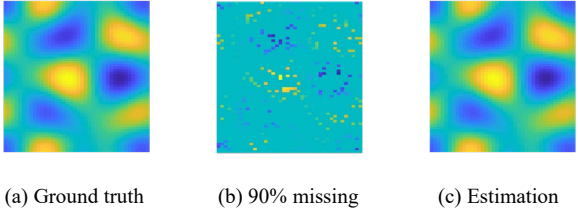

(a) Ground truth    (b) 90% missing    (c) Estimation

Fig. 7. Example of output surfaces, including the actual surface (a), the one with 90% missing values (b), and the estimation using TRMV (c).

### B. Simulation Study for Image Outputs

In this simulation, we evaluate the performance of our proposed method when multiple forms of data are available. Specifically, the waveform surfaces $\mathcal{Y}_i$ are generated based on two input tensors, $\mathcal{X}_{1i} \in \mathbb{R}^{P_{11} \times P_{12} \times \cdots \times P_{1l_1}}$ and $\mathcal{X}_{2i} \in \mathbb{R}^{P_{21} \times P_{22} \times \cdots \times P_{2l_2}}$ $(i = 1, \cdots, m_s)$, where $m_s$ is the sample size. First, $x_{kmj} = \frac{j}{P_{km}}$ $(k = 1,2; m = 1, \cdots, l_k; j = 1, \cdots, P_{km})$ is defined in order to generate the input tensors. Then, we set $\mathbf{U}_{km} = [u_{km1}, u_{km2}, \cdots, u_{kmR_k}]$ $(k = 1,2; m = 1, \cdots, l_k)$, where

$u_{kmt} =$

$\begin{cases} [\cos(2\pi t x_{km1}), \cdots, \cos(2\pi t x_{kmP_{km}})]^T, & \text{if } t \text{ is odd} \\ [\sin(2\pi t x_{km1}), \cdots, \sin(2\pi t x_{kmP_{km}})]^T, & \text{if } t \text{ is even.} \end{cases}$

Then, the elements of a core tensor $\mathcal{D}_{ki}$ are generated randomly from a standard normal distribution. Next, $\mathcal{X}_{ki}$ is simulated by using,

$$\mathcal{X}_{ki} = \mathcal{D}_{ki} \times_1 \mathbf{U}_{k1} \times_2 \cdots \times_{l_k} \mathbf{U}_{kl_k}, (k = 1,2; i = 1, \cdots, m_s).$$

Furthermore, the coefficient tensors $\mathcal{B}_k$ are generated for the regression model. First, a core tensor $\mathcal{C}_{ki}$ is simulated from a standard normal distribution. Next, the basis of output space $\mathbf{V}_m = [v_{m1}, v_{m2}, \cdots, v_{mR}]$ $(m = 1, \cdots, d)$ is generated as,

$v_{mt} =$

$\begin{cases} [\cos(2\pi t y_{m1}), \cdots, \cos(2\pi t y_{mQ_m})]^T, & \text{if } t \text{ is odd} \\ [\sin(2\pi t y_{m1}), \cdots, \sin(2\pi t y_{mQ_m})]^T, & \text{if } t \text{ is even.} \end{cases}$

where $y_{mj} = \frac{j}{Q_m}$. Finally, the coefficient tensors $\mathcal{B}_k$ are computed as,

$$\mathcal{B}_k = \mathcal{C}_k \times_1 \mathbf{U}_{k1} \times_2 \cdots \times_{l_k} \mathbf{U}_{kl_k} \times_{l_k+1} \mathbf{V}_1 \times_{l_k+2} \cdots \times_{l_k+d} \mathbf{U}_d.$$

Given the input tensors and coefficient tensors, the response tensors are calculated as follows,

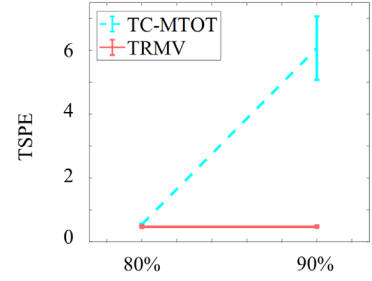

Fig. 8. Performance comparison between the TRMV model and the benchmark under different levels of missing values.

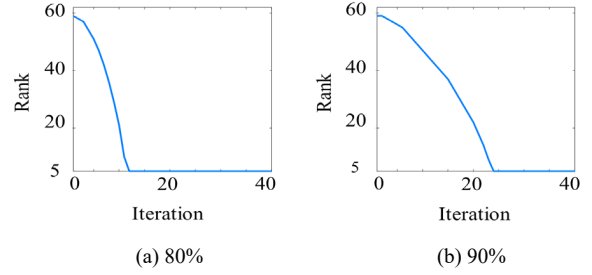

(a) 80%    (b) 90%

Fig. 9. Process of rank estimation of the output tensor at 80% and 90% missing values.

$$\mathcal{Y}_i = \sum_{k=1}^{p} \mathcal{X}_{ki} * \mathcal{B}_k + \mathcal{E}_i,$$

where $\mathcal{E}_i$ is the error tensor whose elements follow a normal distribution $\mathcal{N}(0, \sigma^2)$. We call this data generation steps *procedure B*.

This simulation is conducted for two main purposes: (i) To illustrate the response completion and prediction procedures; (ii) to evaluate the robustness of the proposed method under different noise levels and different rank configurations. First, we generate $p = 2$ inputs with $l_1 = 1$ and $l_2 = 2$, i.e., a profile $\mathcal{X}_{1i} \in \mathbb{R}^{60}$ and an image $\mathcal{X}_{2i} \in \mathbb{R}^{50 \times 50}$ of rank 3. Then, the response tensors $\mathcal{Y}_i \in \mathbb{R}^{60 \times 40}$ are generated with rank 5. Thus, the core tensors of model parameters have dimensions $\mathcal{C}_1 \in \mathbb{R}^{3 \times 5 \times 5}$ and $\mathcal{C}_2 \in \mathbb{R}^{3 \times 3 \times 5 \times 5}$. Here, 200 samples are first generated under $\sigma = 0$, with 100 training samples and 100 testing samples. Then, we randomly remove certain proportions of entries, such as $r = 80\%$ and 90% entries from the training responses. In this study, we also use TSPE to evaluate the performance of the proposed method.

For each level of missing values, we train models using TRMV and TC-MTOT based on the training set. Then, we evaluate and compare the estimated models in terms of their TSPEs, calculated based on the testing set. Fig. 7 illustrates an example of the ground truth surface, the surface with 90% missing values, and its estimation by TRMV. As it is illustrated, the estimation result is highly compatible with the ground truth.

The TSPEs are computed based on the testing set to compare the proposed method to the benchmark. The results are plotted in Fig. 8. Although the TSPE of both methods increases as the percentage of the missing values grows, the magnitudes of TRMV errors are much smaller than those of TC-MTOT at each level. This validates that TRMV has better predictive performance than TC-MTOT.

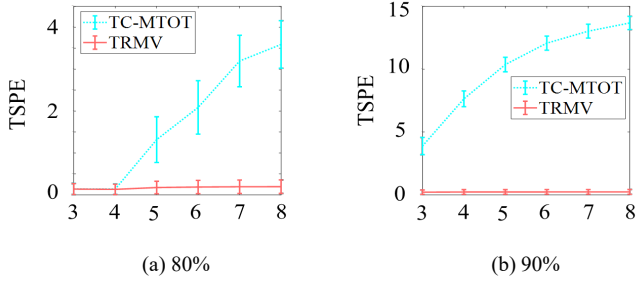

Fig. 10. Performance comparison between TRMV and TC-MTOT at different values of ranks.

TABLE III: PERFORMANCE COMPARISON BETWEEN TRMV AND TC-MTOT AT DIFFERENT LEVELS OF NOISES FOR THE WAVEFORM DATASET

| LMV | 60% | | 70% | | 80% | | 90% | |
|---|---|---|---|---|---|---|---|---|
| $\sigma$ | TRMV | TC-MTOT | TRMV | TC-MTOT | TRMV | TC-MTOT | TRMV | TC-MTOT |
| 0 | **0.1398** | 0.1399 | **0.1400** | 0.1401 | **0.1406** | 0.1407 | **0.2093** | 6.0722 |
| 0.002 | **0.3631** | 0.3684 | **0.3633** | 0.3760 | **0.3635** | 0.4087 | **0.3643** | 6.0727 |
| 0.004 | **0.4079** | 0.4145 | **0.4081** | 0.4240 | **0.4083** | 0.4652 | **0.4093** | 6.0734 |
| 0.006 | **0.4396** | 0.4473 | **0.4398** | 0.4583 | **0.4401** | 0.5058 | **0.4412** | 6.0743 |
| 0.008 | **0.4653** | 0.4739 | **0.4655** | 0.4862 | **0.4658** | 0.5392 | **0.4671** | 6.0753 |
| 0.01 | **0.4874** | 0.4968 | **0.4876** | 0.5102 | **0.4880** | 0.5680 | **0.4893** | 6.0766 |

Furthermore, we evaluate the process of response rank estimation as it is critical for both the response completion and the parameter estimation steps of TRMV. The results are shown in Fig. 9, with the true rank of the response equals 5. Fig. 9 indicates that the proposed algorithm can converge to the true rank in few iterations (around 10 to 30 iterations) in both settings. The result shows the high efficiency of the proposed method.

For the robustness analysis of the proposed method under different ranks and different noise levels, two more datasets with 200 samples are simulated based on *procedure B*. Each setting contains 100 training samples and 100 testing samples. The first dataset is generated with the response $\mathcal{Y}_i \in \mathbb{R}^{20 \times 30}$ of ranks $[3, 4, 5, 6, 7, 8]$ and the inputs $\mathcal{X}_{i1} \in \mathbb{R}^{30}$ and $\mathcal{X}_{i2} \in \mathbb{R}^{20 \times 20}$ of rank 2. Here the noise level is set to be zero (i.e., $\sigma = 0$). Then, we remove entries randomly from the training responses to mimic different levels of missing values. For the second dataset, different levels of noises (i.e., $\sigma = 0.001 \times [0, 2, 4, 6, 8, 10]$) are added to the response. The dimensions of the inputs are $\mathcal{X}_{i1} \in \mathbb{R}^{40}$ and $\mathcal{X}_{i2} \in \mathbb{R}^{40 \times 20}$ of rank 3 and the dimension of the response is $\mathcal{Y}_i \in \mathbb{R}^{20 \times 20}$ of rank 3. Next, we randomly generate an index set $\Omega$ with 90% missing entries and then we project the training set onto $\Omega$, i.e., $\mathcal{P}_{\Omega}(\mathcal{Y})$.

Fig. 10 shows the TSPEs obtained by TRMV and TC-MTOT at different settings of ranks at 80% (panel (a)) and 90% (panel (b)) levels of missing values. In both panels, the dotted lines with error bars show the mean and variance of TSPEs obtained by TC-MTOT; the solid lines with error bars show those obtained by TRMV. Panel (a) of Fig 10 shows that although TC-MTOT achieves comparable TSPEs to TRMV for datasets that are highly correlated and have a simpler structure (i.e., rank is 3 or 4), it performs worse when the structure becomes complicated. This superiority is because of the use of input information in completing the output tensor. From both results illustrated in panels (a) and (b), we conclude that TRMV performs more robustly with smaller mean and variance of prediction errors at every rank setting.

Table III shows the TSPEs obtained by TRMV and TC-MTOT at different levels of noise. As reported in the table, TRMV achieves smaller TSPEs than TC-MTOT at every level of noise and missing values. For example, when $\sigma = 0.006$ and

LMV = 70%, the TRMV achieves TSPE of 0.4398, and TC-MTOT achieves TSPE of 0.4583. Although the TSPEs increase for higher levels of noise and missing values, TRMV performs more robustly in predicting the output than TC-MTOT. This superiority is mainly because TRMV benefits from the systematic and simultaneous estimation of both model parameters and missing values.

## IV. CASE STUDY

In this section, we further evaluate the effectiveness of our proposed method for predicting the overlay (OV) errors in the lithography process. We first introduce the OV errors and then implement TRMV method on an OV dataset.

As shown in Fig. 11 [24], the lithography process aims to transfer a 2-D pattern from an optical mask to a light-sensitive chemical photoresist on the wafer surfaces, which is a critical step in semiconductor manufacturing. During this process, the desired pattern is etched into the material through a series of chemical treatments, such as exposure to light, cleaning, etc. In practice, a wafer may go through such photolithographic cycles for multiple times. During each cycle, one small rectangular field on the wafer surface is completed. Some misalignment error may occur when the patterns on the mask cannot be projected exactly at the desired location on the surface of a wafer. This misalignment is called overlay error and is critical to the wafer quality. As mentioned in the introduction, OV errors are only measured at a limited number of marked locations over the wafer because it is impossible to measure the OV errors of each rectangular field on the wafer. In addition, OV errors are highly dependent on the settings of the lithography machine, which can be used to predict the change of OV errors. The knowledge of OV errors can support the process adjustments, such as the alignments of the projection lens and the location tuning of the wafer stage. To improve the quality of a lithography process, it is critical to correct the OV error by modeling the relationship between the machine settings and the partially measured OV errors. To construct and evaluate our model, we generate a dataset containing the machine settings and overlay error for 200 wafers based on a validated procedure explained in Appendix D. We use this validated simulated data for two main reasons. First, the real data is proprietary. Second, the known "ground truth" OV errors can be used to evaluate the accuracy of our method.

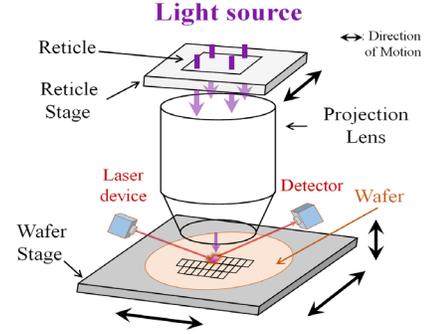

Fig. 11. Illustration of the scanner in semiconductor lithography.

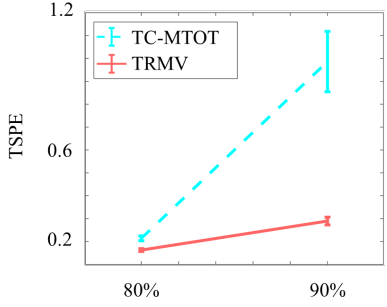

Fig. 12. Performance comparison between TRMV and TC-MTOT under 80% and 90% missing values.

Given the generated dataset, we divide the 200 samples into a training set and a testing set, with 100 samples in each set. For the training set, we assume 80% or 90% points are inaccessible on a wafer. The goal is to generate a model that estimates the OV given the settings of the lithography machine. Given the incomplete training data, we simultaneously estimate the model parameters and complete the missing entries using TRMV. Finally, we investigate the prediction performance of the estimated model in the testing set and calculate the prediction error in terms of the TSPE. In this study, TC-MTOT is also used as the benchmark method for comparison.

Fig. 12 shows the prediction results of two models estimated by TRMV and TC-MTOT at 80% and 90% missing values, respectively. From the results, it is evident that TRMV outperforms TC-MTOT at both levels of missing values.

## V. Conclusion

In this paper, we develop a systematic framework for tensor regression when the response contains missing values. The novelty of this methodology lies in integrating the tensor completion with tensor regression in a unified manner. In order to address the challenge brought by missing values in the response, a penalty of the tensor nuclear norm is introduced into the least square loss function. Meanwhile, Tucker decomposition is applied to the model coefficients to prevent overfitting. A two-step BCD algorithm is proposed to estimate the variables, including the missing entries and the model coefficients. Iteratively, we first complete the missing entries of the response using the ADMM algorithm. Given the completed response, the core tensors and the bases of the coefficients are then estimated by a BCD-ALS algorithm. Notably, the completion procedure of the response in the first step of the algorithm enables the estimation of its rank, which can be used by the second step. These integrated optimization efforts successfully lead to three major advantages of the proposed methodology: (i) more accurate model estimation and response completion; (ii) robust prediction performance; and (iii) automatic rank convergence of the response.

Two simulation studies and a case study are conducted to evaluate the performance of our proposed method in comparison to a two-step method, i.e., TC-MTOT. The numerical results have shown that TRMV outperforms TC-MTOT significantly.

The main contribution of this work is to explore a new direction in tensor regression with missing values. An interesting extension for future study is to consider the selection of input tensors in the sense that some available inputs may not be informative for the response estimation.

## APPENDIX A: Transformation of (7) to (9)

The transformation of the Frobenius norm from problem (7) to problem (9) is derived as follows: Since $\|\mathcal{Y} - \mathcal{A}\|_F^2 = \|\mathbf{Y}_{(i)} - \mathbf{A}_{(i)}\|_F^2, \forall i$ and $\sum_{i=1}^{d} \alpha_i = 1$, we have $\|\mathcal{Y} - \mathcal{A}\|_F^2 = \sum_{i=1}^{d} \alpha_i \|\mathcal{Y} - \mathcal{A}\|_F^2 = \sum_{i=1}^{d} \alpha_i \|\mathbf{Y}_{(i)} - \mathbf{A}_{(i)}\|_F^2$.

## APPENDIX B: Problem (13) transformation

The objective of problem (13) can be transformed as follows: Since $\mathcal{Y}$ and $\Theta_i$ are given, the $\mathcal{M}$-update problem (Equation (14)) is equivalent to,

$$\lambda\alpha_i\|\mathbf{M}_{(i)}\|_* + \frac{\alpha_i}{2}\|\mathbf{M}_{(i)} - \mathbf{A}_{(i)}\|_F^2 - \langle\Theta_i, \mathcal{M}_i\rangle + \frac{\rho}{2}\|\mathcal{Y} - \mathcal{M}_i\|_F^2$$

$$= \lambda\alpha_i\|\mathbf{M}_{(i)}\|_* + \frac{\alpha_i}{2}\|\mathbf{M}_{(i)} - \mathbf{A}_{(i)}\|_F^2$$
$$+ \frac{\rho}{2}\left\|\mathbf{M}_{(i)} - \left(\mathbf{Y}_{(i)} + \frac{1}{\rho}\Theta_{(i)}\right)\right\|_F^2$$

$$= \lambda\alpha_i\|\mathbf{M}_{(i)}\|_* + \frac{\alpha_i+\rho}{2}\left\|\mathbf{M}_{(i)} - \frac{\alpha_i\mathbf{A}_{(i)}+\rho\mathbf{C}_{0(i)}}{\alpha_i+\rho}\right\|_F^2 + b, \quad \text{(B1)}$$

where $b$ is a constrant and $\mathbf{C}_{0(i)} = \left(\mathbf{Y}_{(i)} + \frac{1}{\rho}\Theta_{(i)}\right)$.

## APPENDIX C: Solution to (17)

In this appendix, we derive the solution for problem (17). Let the partial derivative of Lagrangian $L_y(\mathcal{Y}, \phi)$ be equal to zero and we have,

$$\begin{cases} \frac{\partial L_y(\mathcal{Y},\phi)}{\partial y} = \sum_{i=1}^{d}\left\{\Theta_i + \frac{\rho}{2}(\mathcal{Y} - \mathcal{M}_i)\right\} + \mathcal{P}_\Omega(\phi) = 0 \\ \frac{\partial L_y(\mathcal{Y},\phi)}{\partial \phi} = \mathcal{P}_\Omega(\mathcal{Y} - \mathcal{Y}_0) = 0 \end{cases} \quad \text{(C1)}$$

When $(j, q_1, \cdots, q_d) \notin \Omega$, we only have,

$$\sum_{i=1}^{d}\left\{\Theta_i + \frac{\rho}{2}\left(\mathcal{Y}_{(j,q_1,\cdots,q_d)} - \mathcal{M}_i\right)\right\} = 0. \quad \text{(C2)}$$

Otherwise, we have,

$$\begin{cases} \mathcal{P}_\Omega(\Phi) = 0 \\ \mathcal{P}_{\bar{\Omega}}(\mathcal{Y} - \mathcal{Y}_0) = 0 \end{cases}. \tag{C3}$$

Thus, we can obtain the optimal solution of the minimization problem as,

$$\mathcal{Y} = \begin{cases} \mathcal{Y}_0, \ if \ (j, q_1, \cdots, q_d) \in \Omega \\ \frac{1}{d}\sum_{i=1}^{d}\left(\mathcal{M}_i - \frac{1}{\rho}\Theta_i\right), otherwise \end{cases}. \tag{C4}$$

## APPENDIX D: CASE STUDY DATA GENERATION

This appendix elaborates on the popular procedure used for generating the OV errors in the lithography process. The OV error is denoted as a vector $(F_x, F_y)$ in 2D coordinate system, with the initial point $(x, y)$ and the terminal point $(x + F_x, y + F_y)$ representing the locations of the previous layer and the current layer, respectively. To maintain OV requirement, the polynomial model is widely used in practice for overlay control [25], [26], which is represented as follows,

$$\begin{cases} F_x(x, y) = \mathbf{k}_x\mathbf{b} \\ F_y(x, y) = \mathbf{k}_y\mathbf{b} \end{cases}, \tag{D1}$$

where $F_x$ and $F_y$ are the coordinates of the OV error; $\mathbf{b} = [1, x, y, x^2, xy, y^2, x^3, x^2y, xy^2, y^3]^T \in \mathbb{R}^{10}$ is a basis; $\mathbf{k}_x = [k_1, k_3, k_5, ..., k_{19}]$ and $\mathbf{k}_y = [k_2, k_4, k_6, ..., k_{20}]$ define the signatures of an OV error, which are used for on-line correction via machine settings such as wafer stage, lens, mask, and so on, along the motion directions. With this model, the OV error can be broken into linear components, i.e., $(k_1, k_2, ..., k_6)$ and nonlinear components, i.e., $(k_7, k_8, ..., k_{20})$, where the linear ones associate with errors including reticle rotation, wafer rotation, to name a few, and the nonlinear ones correspond to causes such as lens distortion and random errors [26].

Specifically, a set of points in a wafer are first specified to construct a basis $\mathbf{B} = [\mathbf{b}_1, \mathbf{b}_2, ..., \mathbf{b}_n]$. For the signature, we only take the variation of linear components into consideration, which means $(k_7, k_8, ..., k_{20})$ equals to zeros. The variation patterns of the signature $(k_1, k_2, ..., k_6)$ are randomly generated and we obtain $\mathbf{K}_x^i = [k_1^i, k_3^i, k_5^i, 0, ..., 0]$ and $\mathbf{K}_y^i = [k_2^i, k_4^i, k_6^i, 0, ..., 0]$ for $i = 1, ..., m$. Then, for each $(\mathbf{K}_x^i, \mathbf{K}_y^i)$, the associated OV pattern $(\mathbf{F}_x^i, \mathbf{F}_y^i)$ can be calculated as $\mathbf{F}_x^i = [F_x^{i1}, ..., F_x^{in}]$ and $\mathbf{F}_x^i = [F_y^{i1}, ..., F_y^{in}]$.

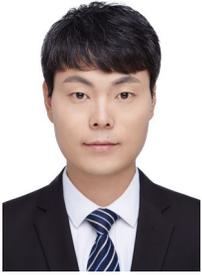

**Feng Wang** (Student Member, IEEE) received the M.S. degree in Traffic Information Engineering & Control from Beijing Jiaotong University, Beijing, China, in 2016, where he is working toward the Ph.D. degree with the State Key Laboratory of Rail Traffic Control and Safety. Currently, he is a visiting scholar in the H. Milton Stewart School of Industrial and Systems Engineering, Georgia Institute of Technology, Atlanta. His research interests include statistical modeling and prognostics with high-dimensional data.

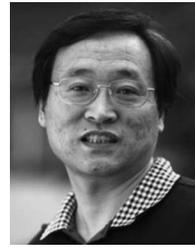

**Tao Tang** (Senior Member, IEEE) received the Ph.D. degree in automatic control theory and application from Chinese Academy of Sciences, Beijing, China, in 1991.

He is currently the Director of the School of Electronic and Information Engineering and the State Key Laboratory of Rail Traffic Control and Safety, Beijing Jiaotong University, Beijing, China. His research interests include both high-speed and urban railway train control systems, as well as intelligent transportation systems. He is a member of the Experts Group of High Technology Research and Development Program of China (863 Program) and the Leader in the Field of the Modern Transportation Technology Experts Group. He is also a Specialist in the National Development and Reform Commission and the Beijing Urban Traffic Construction Committee.

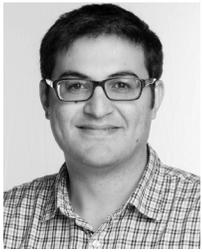

**Mostafa Reisi-Gahrooei** received the master's degree in computational science and engineering and the Ph.D. degree in industrial and systems engineering from Georgia Institute of Technology, Atlanta, GA, USA, and the M.Sc. degrees in transportation engineering and applied mathematics from the Southern Illinois University Edwardsville, Edwardsville, IL, USA.

He is currently an Assistant Professor with the Department of Industrial and Systems Engineering, the University of Florida, Gainesville, FL, USA. His research focuses on modeling, monitoring, and control of complex systems with functional, high-dimensional data.

Dr. Reisi Gahrooei is a member of the Institute for Operations Research and the Management Sciences (INFORMS) and the Institute of Industrial and Systems Engineers (IISE).

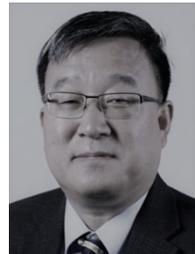

**Jianjun Shi** received the B.S. and M.S. degrees in electrical engineering from the Beijing Institute of Technology, Beijing, China, in 1984 and 1987, respectively, and the Ph.D. degree in mechanical engineering from the University of Michigan, Ann Arbor, in 1992.

He is the Carolyn J. Stewart Chair and Professor in H. Milton Stewart School of Industrial and Systems Engineering, with joint appointment in the George W. Woodruff School of Mechanical Engineering, Georgia Institute of Technology. Dr. Shi's research interests are in the development and application of data enabled manufacturing. His methodologies integrate system informatics, advanced statistics, and control theory for the design and operational improvements of manufacturing and service systems by fusing engineering systems models with data science methods.

Dr. Shi is a Fellow of IIE, ASME, and INFORMS, an Academician of the International Academy for Quality, and a member of U.S. National Academy of Engineering (NAE).

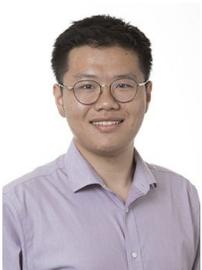

**Zhen Zhong** received the B.S. degree in Electrical Engineering from University of Science and Technology of China, Anhui, China, in 2017.

Currently, he is a Ph.D. student at the H. Milton Stewart School of Industrial and Systems Engineering, Georgia Institute of Technology, Atlanta. His research interests are focused on the process control in semiconductor manufacturing.